\documentclass[a4paper]{svmult}
\usepackage{times}
\usepackage[dvips]{graphicx,epsfig}
\usepackage{amsmath,amsfonts,amssymb}
\usepackage{cite,url}

\begin{document}



\mainmatter

\title*{Proton Medium Modifications from $^4$He$(\vec e,e' \vec p)^3$H}

\author{\underline{{S.~Strauch}}
\and {S.~Malace}
\and {M.~Paolone}
for the Jefferson Lab Hall A Collaboration
}

\titlerunning{Proton Medium Modifications from $^4$He$(\vec e,e' \vec p)^3$H}
\authorrunning{{S.~Strauch}, {S.~Malace}, and {M.~Paolone}}

\toctitle{Proton Medium Modifications from $^4$He$(\vec e,e' \vec p)^3$H}
\tocauthor{{S.~Strauch}, {S.~Malace}, {M.~Paolone}}

\institute{{University of South Carolina, Columbia, South Carolina 29208, USA}}

\maketitle

\begin{abstract} Polarization transfer in quasi-elastic nucleon knockout is sensitive to the properties of the nucleon in the nuclear medium.  In experiment E03-104 at Jefferson Lab we measured the proton recoil polarization in the $^4$He($\vec e,e^\prime \vec p\,$)$^3$H reaction at a $Q^2$ of 0.8 (GeV/$c$)$^2$ and 1.3 (GeV/$c$)$^2$ with unprecedented precision. The measured polarization-transfer coefficients transverse and longitudinal to the momentum-transfer direction are well described by a fully relativistic calculation when a density-dependent medium modification of the nucleon form factors is included in the model. Results of an alternative model show that the ratio of these observables is also well described through strong charge-exchange final-state interactions. The induced polarization in the $(e,e'\vec p)$ reaction is sensitive to the final-state interactions and the data from E03-104 will further constrain these models.  \end{abstract}

\section{Introduction}

Quantum chromodynamics (QCD) is established as the theory of the strong nuclear force but the degrees of freedom observed in nature, hadrons and nuclei, are different from those appearing in the QCD Lagrangian, quarks and gluons. There are no calculations available for nuclei within the QCD framework. Nuclei are effectively and well described as clusters of protons and neutrons held together by a strong, long-range force mediated by meson exchange \cite{mosz60}. Whether the nucleon changes its internal structure while embedded into a nuclear medium has been a long-standing question in nuclear physics \cite{Saito07}. The nuclear European Muon Collaboration (EMC) effect seems to suggest the modification of hadrons in the nuclear medium; see \cite{Smith04}. The issue has attracted theoretical attention and various models have been developed to study medium modifications. A list of some of these models includes the many-body soliton dynamics model by Celenza {\it et al.} \cite{Celenza85}, the quark-meson coupling (QMC) model by Lu {\it et al.}~\cite{Lu98}, the light-front-constituent quark model by Frank {\it et al.} \cite{Fr96}, the modified Skyrme model by Yakshiev {\it et al.}~\cite{Ya02}, the chiral quark-soliton model (CQS) by Smith and Miller \cite{Smith04}, and the Nambu-Jona-Lasinio model of Horikawa and Bentz \cite{Horikawa05}. The connection between in-medium modifications of the nucleon form factors and of the deep inelastic structure functions is discussed by Liuti \cite{Liuti06} using the concept of generalized parton distributions (GPDs). Guzey {\it et al.} \cite{guzev} have studied incoherent deeply virtual Compton scattering on $^4$He in the $^4{\rm He}(e,e^{\prime}\gamma p)X$ reaction, which probes medium-modifications of the bound nucleon GPDs and elastic form factors.  Medium modifications of nucleon properties in nuclear matter and finite nuclei have been also discussed by Wen and Shen \cite{Wen08}.

The question has also attracted experimental attention. One of the most intuitive methods to investigate the properties of nucleons inside nuclei is quasi-elastic scattering off nuclei. Since the charge and magnetic responses of a single nucleon are quite well studied from elastic scattering experiments, measuring the same response from quasi-elastic scattering off nuclei and comparing with a single nucleon is likely to shed light on the question.  The ratio of polarization-transfer coefficients in elastic $\vec e p$ scattering, $P'_x/P'_z$, is directly proportional to the ratio of the electric and magnetic form factors of the proton, $G_E/G_M$ \cite{Arnold81},
\begin{equation}
 \frac{P'_x}{P'_z} = -\frac{G_E}{G_M}\frac{2m}{E_i+E_f}\tan^{-1}\frac{\theta_e}{2};
\end{equation}
here $P'_x$ and $P'_z$ are the polarization-transfer coefficients transverse and longitudinal to the momentum-transfer direction,
\begin{eqnarray}
P'_x &=& -2\sqrt{\tau(1+\tau)}\frac{\frac{G_E}{G_M}}{(\frac{G_E}{G_M})^2+\frac{\tau}{\epsilon}}\tan\frac{\theta_e}{2}, \label{eq:px}\\
P'_z &=& \frac{1}{m} \left(E_i+E_f\right) \sqrt{\tau(1+\tau)}\frac{1}{(\frac{G_E}{G_M})^2+\frac{\tau}{\epsilon}}\tan^2\frac{\theta_e}{2}, \label{eq:pz}
\end{eqnarray}
and $\epsilon = \left[1 + 2(1+\tau)\tan^2\frac{\theta_e}{2} \right]^{-1}$ and $\tau=Q^2/4m^2$ are kinematic variables, $E_i$ and $E_f$ are the incident and final electron energies, $\theta_e$ is the electron scattering angle, $m$ is the nucleon mass, and $Q^2$ is the four-momentum transfer squared.

When such measurements are performed on a nuclear target in quasi-elastic kinematics, the polarization-transfer observables are sensitive to the form-factor ratio of the proton embedded in the nuclear medium.  However, distinguishing possible changes in the structure of nucleons embedded in a nucleus from more conventional many-body effects like meson-exchange currents (MEC), isobar configurations (IC) or final-state interactions (FSI) is only possible within the context of a model. Experimental results for the polarization-transfer ratio are conveniently expressed in terms of the polarization double ratio,
\begin{equation}
R = \frac{(P_x'/P_z')_{A}}{(P_x'/P_z')_{^1\rm H}},
\label{eq:rexp}
\end{equation}
in order to emphasize differences between the in-medium compared to the free values. Here, the polarization-transfer ratio for the quasi-elastic proton knockout reaction, $A(\vec e, e'\vec p)$, is normalized to the polarization-transfer ratio measured off hydrogen in the identical setting. Such a double ratio cancels nearly all experimental systematic uncertainties.

\section{Experiments}

Experiment E89-033 was the first to measure the polarization transfer $(\vec e,e' \vec p)$ in a complex nucleus, $^{16}$O \cite{Malov00}.  The results are consistent with predictions of relativistic calculations based on the free proton from factor within a rather large experimental uncertainty of about 18\%. Polarization-transfer experiments have studied nuclear medium effects in deuterium \cite{Eyl95,Mil98,Bark99} at the Mainz microtron (MAMI) and MIT-Bates and more recently at 
the Thomas Jefferson National Accelerator Facility (JLab) \cite{Hu06}. Within statistical uncertainties, no evidence of medium modifications was found and the data are well described by a model calculation of Arenh\"ovel, which includes final-state interactions (FSI), meson-exchange currents (MEC), and isobar configurations, as well as relativistic contributions; see \cite{Hu06}. As the sampled density is small, it is not surprising that there are no indications for medium modifications of the proton electromagnetic form factors in the $^2$H data.

One might expect to find larger medium effects in $^4$He, with its significantly higher density.  Indeed, a recent JLab Experiment has made a precision measurement of the EMC effect in both few-body nuclei and heavy nuclei. The findings indicate that the nuclear dependence of the deep-inelastic cross section for $^4$He is comparable to that for $^{12}$C \cite{Seely09}.  Although estimates of the many-body effects in $^4$He may be more difficult than in $^2$H, calculations for $^4$He indicate they are small \cite{Laget94}. The first $^4$He$(\vec e,e^\prime \vec p)^3$H polarization-transfer measurements were performed at MAMI at $Q^2 = 0.4$ (GeV/$c$)$^2$ \cite{Dieterich01} and at Jefferson Lab Hall A at $Q^2$ = 0.5, 1.0, 1.6, and 2.6 (GeV/$c$)$^2$ \cite{Strauch03}. Our recent experiment E03-104 \cite{PR03104} extended these measurements with two high-precision points at $Q^2$ = 0.8 and 1.3 (GeV/$c$)$^2$.  All these data were taken in quasi-elastic kinematics at low missing momentum with symmetry about the three-momentum-transfer direction to minimize conventional many-body effects in the reaction. E03-104 covers a range of missing momenta up to about 135 MeV/$c$. In these experiments, two high-resolution spectrometers were used to detect the scattered electron and the recoil proton in coincidence. The missing-mass technique was used to identify $^3$H in the final state.  The proton spectrometer was equipped with a focal plane polarimeter (FPP).  Polarized protons lead to azimuthal asymmetries after scattering in the carbon analyzer of the FPP. These distributions, in combination with information on the spin precession of the proton in the magnetic fields of the spectrometer, the carbon analyzing power, and the beam helicity, were analyzed by means of a maximum likelihood method to obtain the induced polarization, $P_y$, and polarization transfer components, $P'_x$ and $P'_z$ \cite{Punjabi05}. As the experiment was designed to detect differences between the in-medium polarizations and the free values, both $^4$He and $^1$H targets were employed.

In the polarization-transfer double ratio $R$, nearly all systematic uncertainties cancel: the polarization-transfer observables are not sensitive, to first order, to the instrumental asymmetries in the FPP, and their ratio is independent of the electron beam polarization and the graphite analyzing power. The small systematic uncertainties are due, mainly, to the uncertainties in the spin transport through the proton spectrometer but an extensive study is being performed in order to reduce their contribution to the total uncertainty. The induced proton polarization $P_{y}$ is a direct measure of final-state interactions. However, the $P_{y}$ extraction is greatly complicated by the presence of instrumental asymmetries in the FPP. Typically, instrumental asymmetries are due to detector misalignments, detector inefficiencies or tracking issues. An ongoing effort aimed at devising a method to minimize instrumental asymmetries will make possible the precise extraction of the induced polarization $P_{y}$ from E03-104 measurements.

\section{Results}

\begin{figure}[h]
  \includegraphics[width=6cm]{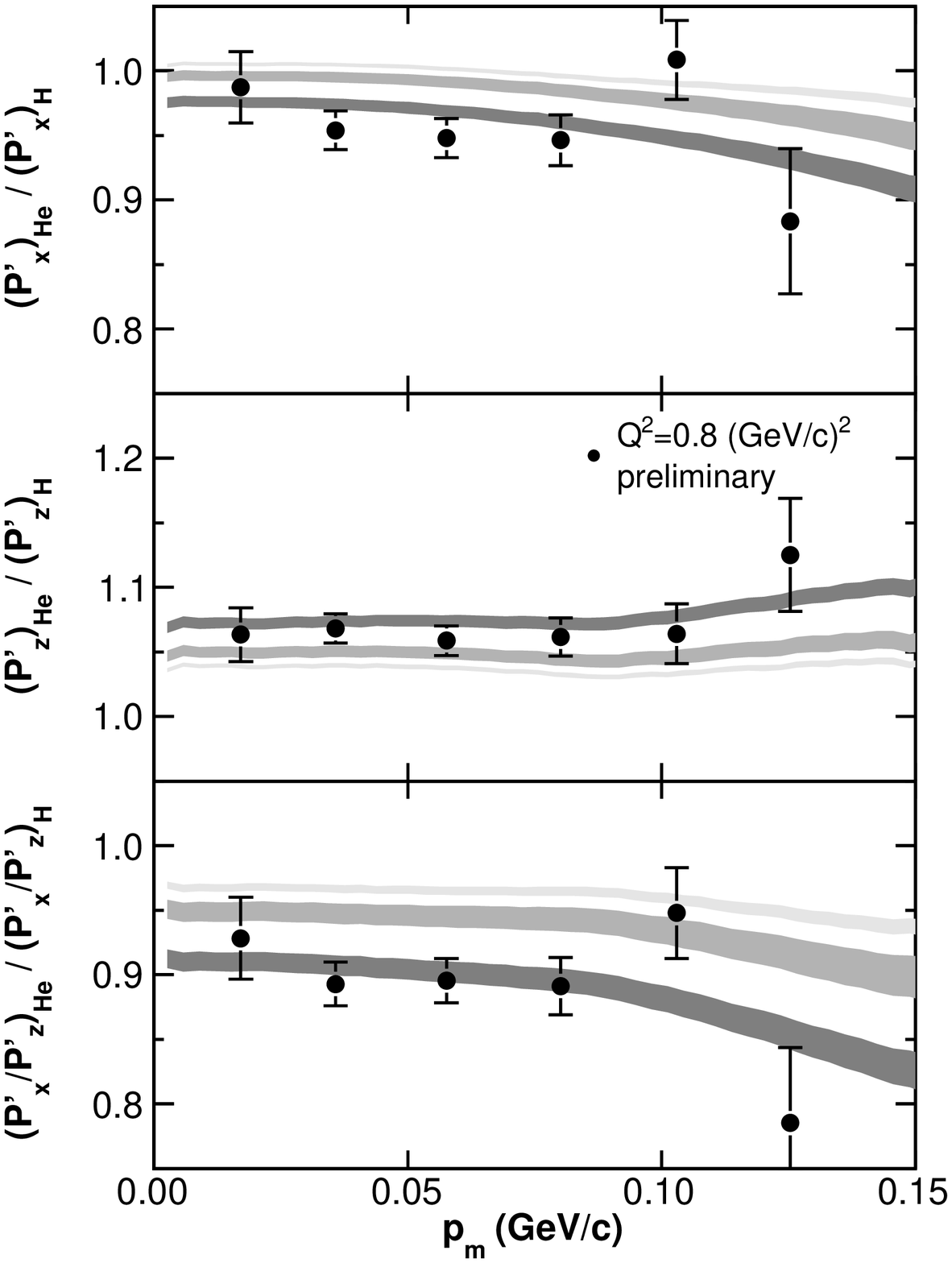}\hfill
  \includegraphics[width=6cm]{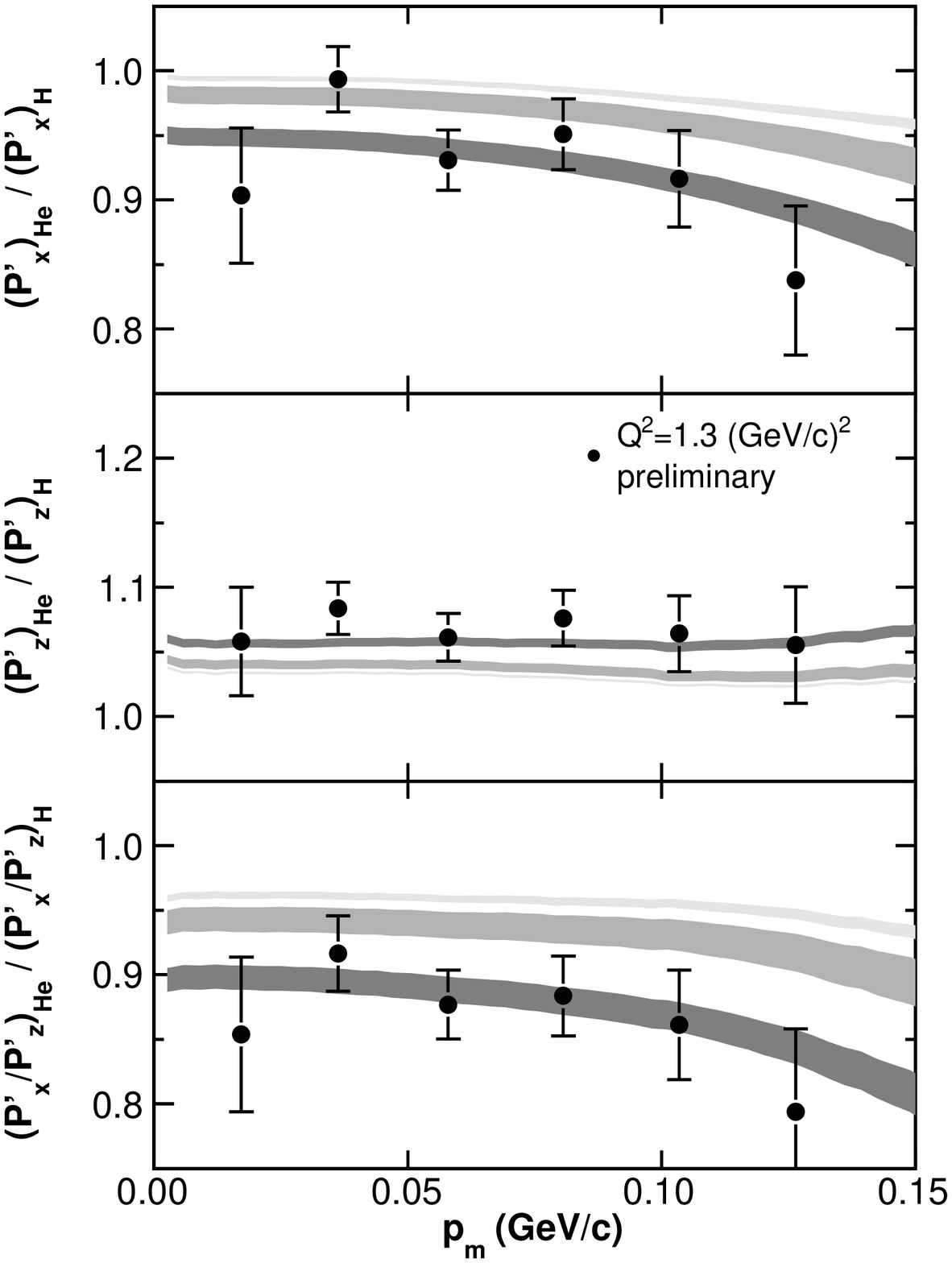}
  \caption{Preliminary {$^4$He$(\vec e, e'\vec p)^3$H
      polarization-transfer data from E03-104 at $Q^{2}$ of 0.8 and 1.3 (GeV/$c$)$^{2}$.
      The data are normalized to the respective experimental results from the elastic $^1$H$(\vec e,e' \vec p)$ reaction. The bands represent results of RPWIA (light band) and RDWIA calculations from the Madrid group \cite{Ud98} including, respectively,  free nucleon form factors (medium band) and medium-modified form factors from the QMC model \cite{Lu98} (dark band) in the calculations.
      \label{fig:pxpzr}} }
\end{figure}

The preliminary $^4$He$(\vec e, e' \vec p)^3$H polarization-transfer observables $P'_x$ and $P'_z$ and their ratios $P'_x/P'_z$ are shown in Figure \ref{fig:pxpzr} for various missing momentum bins and at $Q^{2}$ of 0.8 (left panel) and 1.3 (GeV/$c$)$^{2}$ (right panel). The data are normalized by the experimental results from elastic $^1$H$(\vec e, e' \vec p)$ scattering. These ratios show most clearly the difference between quasi-elastic and elastic results. An experimental advantage of these ratios is additionally the cancellation of the carbon analyzing power. The data show that the $^4$He results are systematically low compared to the $^1$H results for $P'_x$ and high for $P'_z$ resulting in an about 10\% to 12\% quenching of the polarization-transfer ratio $P'_x/P'_z$. The amount of quenching in the data appears to increase with missing momentum.

The data are compared with model calculations by the Madrid group \cite{Ud98} based on a relativistic plane-wave impulse approximation (RPWIA), a relativistic distorted-wave impulse approximation (RDWIA), and the RDWIA including medium-modified nucleon form factors from the QMC model \cite{Lu98} (RDWIA+QMC). Results of these calculations are shown in Figure \ref{fig:pxpzr} as light, medium, and dark bands, respectively. The widths of the bands indicate the variation in the results of the calculations using various input: the $cc1$ or $cc2$ current operators as defined in \cite{Forest83} and the relativistic optical potentials by McNeil, Ray, and Wallace\cite{McNeil83} (MRW) or Horowitz \cite{Horowitz} (RLF) to model FSI. All these calculations use the Coulomb gauge. MEC are not explicitly included in the Madrid calculation. Predictions by Meucci {\it et al.} \cite{meucci} show that the two-body current (the seagull diagram) effects are generally small (less than 3 \% close to zero missing momenta) and visible only at high missing momenta.  

The sets of calculations give distinctively different results even with these changes in the model input and the small uncertainties from E03-104 allow to discriminate between the various sets of calculations. Both, RPWIA and RDWIA calculations do not describe the data; this is most noticeable in the polarization-transfer double ratio, $R$, where RDWIA accounts for only about half of the observed quenching. The data favor the inclusion of the density-dependent in-medium form factors from the QMC model into the RDWIA calculations in all cases. The inclusion of the in-medium form factors leads to a relative reduction of $P'_x$ and a relative increase of $P'_z$ compared to the RDWIA results with a resulting effect of lowering the polarization-transfer ratio by an additional 5\% to 6\% and bringing it into agreement with data.  It is interesting to note that main differences in the results of the RDWIA and RDWIA+QMC calculations can be quantitatively understood by naively applying the elastic $\vec e p$ scattering formalism to the quasielastic case.  From Eq. (\ref{eq:rexp}) one would expect a reduction in $P'_x/P'_z$ from a decreased, in-medium value of $G_E/G_M$. From Eqs. (\ref{eq:px}) and (\ref{eq:pz}) we can then estimate the changes of the individual polarization-transfer coefficients due to a variation of $G_E/G_M$; these depend on the kinematics of the experiment, $\tau/\epsilon$. With a decrease of $P'_x/P'_z$ by about 5\% this leads for the particular kinematics of our experiment at $Q^2$ of 0.8 (GeV/$c$)$^2$ to a relative decrease of $P'_x$ by about 2\% and to a relative increase of $P'_z$ by about 3\%. Due to its different $\tau/\epsilon$ ratio, these relative changes are, respectively,  3.5\% and 1.5\% at $Q^2$ of 1.3 (GeV/$c$)$^2$; in agreement with the full model.

\begin{figure}[htb!]
  \begin{center}
    \includegraphics[width=\textwidth]{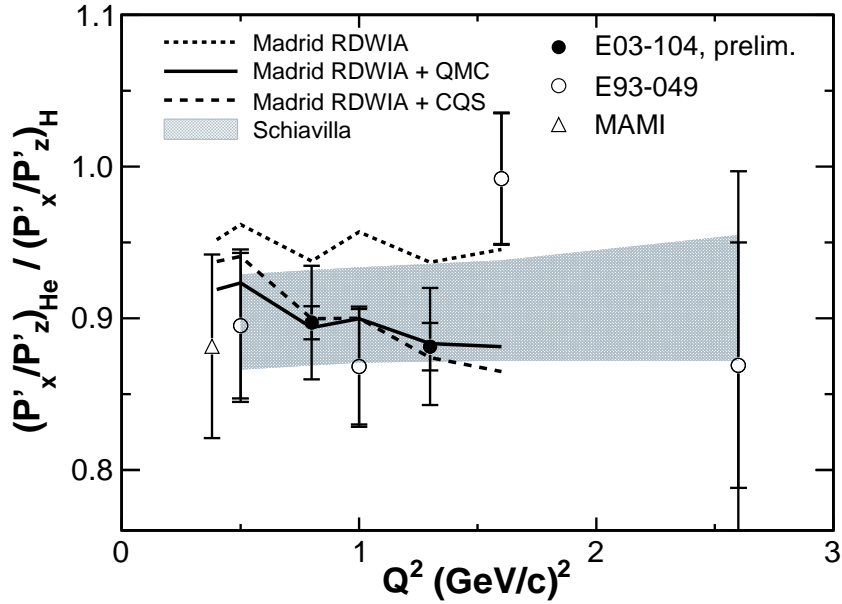}
    \caption{$^4$He$(\vec e, e'\vec p)^3$H
      polarization-transfer double ratio $R$ as a function of
      $Q^2$ from Mainz \cite{Dieterich01} and Jefferson Lab experiment
      E93-049 \cite{Strauch03} (open symbols) along with preliminary
      results from experiment E03-104 (filled circles).  The data are
      compared to calculations from the Madrid group \cite{Ud98} and
      Schiavilla {\it et al.} \cite{Schiavilla05}. In-medium form
      factors from the QMC \cite{Lu98} (solid curve) and CQS
      \cite{Smith04} (dashed curve) models were used in two of the
      Madrid calculations.
      Not shown are a relativistic Glauber model calculation by the
      Ghent group \cite{Lava05} and results from Laget \cite{Laget94} which both give
      a value of $R\approx 1$.
      \label{fig:ratio}}
  \end{center}
\end{figure}

The $^4$He polarization-transfer double ratio is shown in Figure \ref{fig:ratio} for all available data. The preliminary data from E03-104 (filled circles) are consistent with the previous data from JLab E93-049 \cite{Strauch03} and MAMI \cite{Dieterich01} (open symbols). The polarization-transfer ratios $P'_x/P'_z$ in the $(\vec e,e' \vec p)$ reaction on helium are significantly different from those on hydrogen.  The data are compared with results of the RDWIA calculation by the Madrid group \cite{Ud98} (dotted curve).  The calculation shown uses the Coulomb gauge, the $cc1$ current operator as defined in \cite{Forest83}, and the MRW optical potential of \cite{McNeil83}.  The $cc2$ current operator gives higher values of $R$, worsening agreement with the data.  It can be seen that the Madrid RDWIA calculation (dotted curve) overpredicts the data by about 6 \%.  We note that these relativistic calculations provide good descriptions of, e.g., the induced polarizations as measured at Bates in the $^{12}$C(e,e$^\prime \vec p$) reaction \cite{Woo98} and of $A_{TL}$ in $^{16}$O($e, e^\prime p$) as previously measured at JLab \cite{Gao00}.  After including the density-dependent medium-modified form factors from the QMC \cite{Lu98} or CQS \cite{Smith04} models in the RDWIA calculation (solid and dashed curves), good agreement with the polarization-transfer data is obtained. As the proton momentum at $Q^2 = 2.6$ (GeV/$c$)$^2$ exceeds the range of validity of the MRW optical potential no calculations by the Madrid group are shown beyond 1.6 (GeV/$c$)$^2$.

This agreement has been interpreted as possible evidence of proton medium modifications \cite{Strauch03}. It is based on the excellent description of the data by a particular model in terms of medium modifications of nucleon form factors and requires good control of the reaction mechanisms such as meson-exchange currents, isobar configurations, and final-state interactions.  In fact, there is an alternative interpretation of the observed suppression of the polarization-transfer ratio within a more traditional calculation by Schiavilla {\it et al.} \cite{Schiavilla05} (shaded band). Schiavilla's calculation uses free nucleon form factors. Explicitly included MEC effects paired with tensor correlations suppress $R$ by almost 4\% in his calculation.  The FSIs are treated within the optical potentials framework and include both a spin-dependent and spin-independent charge-exchange term; the spin-orbit terms, however, are not well constrained by data.  In Schiavilla's model, the final-state interaction effects suppress $R$ by an additional 6\% bringing this calculation also in good agreement with the data within the statistical uncertainties associated with the Monte Carlo technique in this calculation. It should be noted that charge-exchange terms are not taken into account in the Madrid RDWIA calculation. The difference in the modeling of final-state interactions is the origin of the major part of the difference between the results of the calculations by the Madrid group \cite{Ud98} and Schiavilla {\it et al.}~\cite{Schiavilla05} for the polarization observables.

\begin{figure}
\begin{center}
  \includegraphics[width=\textwidth]{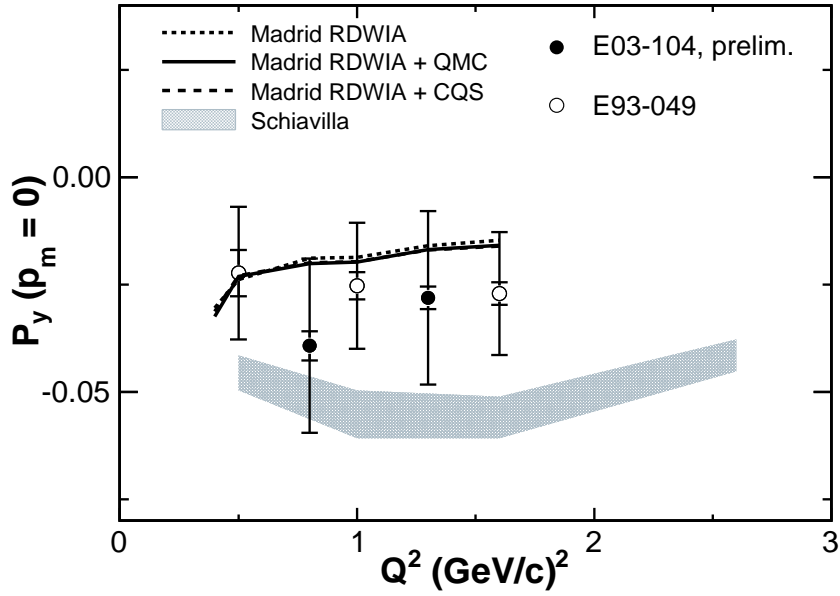}
  \caption{Induced polarization data from
    Jefferson Lab experiment E93-049 \cite{Strauch03} along with preliminary
    results from experiment E03-104. The data are compared to
    calculations from the Madrid group \cite{Ud98} and Schiavilla {\it et al.}
    \cite{Schiavilla05}. The comparison is made for missing momentum
    $p_m \approx 0$; note that the experimental data have been
    corrected for the spectrometer acceptance for this comparison.  }
  \label{fig:py}
\end{center}
\end{figure}

Effects from final-state interactions can be studied experimentally with the induced polarization, $P_y$, which vanishes in the absence of final-state interactions. Induced-polarization data were taken simultaneously to the polarization-transfer data.  Figure \ref{fig:py} shows the preliminary data for $P_y$. The induced polarization is small at the low missing momenta in this reaction. The sizable systematic uncertainties are due to possible instrumental asymmetries.  Dedicated data have been taken during E03-104 to study these and work is underway to significantly reduce the systematic uncertainties in $P_y$ in the final analysis. The data are compared with the results of the calculations from the Madrid group and Schiavilla {\it et al.}  at missing momenta of about zero. To facilitate this comparison, the data have been corrected for the spectrometer acceptance.
The preliminary data suggest that the measured induced polarization (and thus the final-state interaction) is overestimated in the model of Schiavilla {\it et al.} Note that the charge-exchange terms, particularly, the spin-dependent one, gives the largest contribution to Schiavilla's calculation of P$_{y}$.  The induced polarization proves to be sensitive to the choice of optical potential allowing this observable to be used to constrain theoretical models of FSI.

A comparison of the model calculations in Figure \ref{fig:ratio} and Figure \ref{fig:py} shows that the in-medium form factors mostly affect the ratio of polarization-transfer observables, not the induced polarization. It is a great advantage of E03-104 to have access to both the polarization-transfer and the induced polarization.

In summary, polarization transfer in the quasi-elastic $(e,e^\prime p)$ reaction is sensitive to possible medium modifications of the bound-nucleon form factor. Currently, the $^4$He$(\vec e, e^\prime \vec p)^3$H polarization-transfer data can be well described by either the inclusion of medium-modified form factors or strong charge-exchange FSI in the models. However, these strong FSI effects may not be consistent with the induced polarization data. The final analysis of the new high-precision data from Jefferson Lab Hall A should provide a more stringent test of these calculations.

\section*{Acknowledgments}

This work was supported in parts by the U.S. National Science Foundation: NSF PHY-0856010.
Jefferson Science Associates operates the Thomas Jefferson National Accelerator Facility under DOE
contract DE-AC05-06OR23177.


\begin{thebibliography}{32}
\bibitem{mosz60} S.A. Moszkowski and B.L. Scott, \emph{ Ann. Phys. (Paris)} {\bf 11}, 65 (1960).
\bibitem{Saito07} K.~Saito, K.~Tsushima and A.~W.~Thomas, \emph{ Prog.\ Part.\ Nucl.\ Phys.\ }  {\bf 58}, 1 (2007).
\bibitem{Smith04} J.R.~Smith and G.A.~Miller, \emph{ Phys. Rev. C}  {\bf 70}, 065205 (2004).
\bibitem{Celenza85} L.S.~Celenza, A.~Rosenthal, and C.M. Shakin, \emph{ Phys.~Rev.~C} {\bf 31}, 232 (1985).
\bibitem{Lu98} D.H.~Lu \emph{ et al.}, \emph{ Phys.~Lett. B} {\bf 417}, 217 (1998);
\emph{ Phys.~Rev. C} {\bf 60}, 068201 (1999).
\bibitem{Fr96} M.R.~Frank, B.K.~Jennings, and G.A.~Miller, \emph{ Phys.~Rev. C} {\bf 54}, 920 (1996).
\bibitem{Ya02} U.T.~Yakhshiev, U-G. Meissner, and A. Wirzba,
\emph{ Eur.~Phys.~J.} A {\bf 16}, 569 (2003).
\bibitem{Horikawa05} T.~Horikawa and W.~Bentz, \emph{ Nucl.~Phys.~A} {\bf 762}, 102 (2005).
\bibitem{Liuti06} S. Liuti, arXiv:hep-ph/0601125v2.
\bibitem{guzev} V. Guzey, A.W. Thomas, K. Tsushima, \emph{ Phys. Lett. B} {\bf 673}, 9 (2009).
\bibitem{Wen08} W.~Wen and H.~Shen, \emph{ Phys.~Rev.~C} {\bf 77}, 065204 (2008).
\bibitem{Arnold81} R.G. Arnold, C.E. Carlson, and F.~Gross, \emph{ Phys.~Rev.~C} {\bf 23}, 363 (1981).
\bibitem{Malov00} S.~Malov \emph{ et al.}, (The Jefferson Laboratory Hall A
Collaboration) \emph{ Phys.~Rev.~C} {\bf 62}, 057302 (2000).
\bibitem{Eyl95} D. Eyl \emph{ et al.}, \emph{ Z. Phys. A} {\bf 352}, 211 (1995).
\bibitem{Mil98} B.D. Milbrath, J. McIntyre \emph{ et al.}, \emph{ Phys.~Rev.~Lett.}. {\bf 80}, 452 (1998).
\bibitem{Bark99}D.H. Barkhuff, \emph{ et al.}, \emph{ Phys.~Lett.~B} {\bf 470}, 39 (1999).
\bibitem{Hu06} B. Hu \emph{ et al.}, \emph{ Phys.~Rev.~C} {\bf 73}, 064004 (2006).
\bibitem{Seely09} J.~Seely \emph{ et al.}, \rm arXiv:0904.4448v2 [nucl-ex] 5 May (2009).
\bibitem{Laget94} J.M.~Laget, \emph{ Nucl.~Phys.~A} {\bf 579}, 333 (1994). %
\bibitem{Dieterich01} S.~Dieterich \emph{ et al.}, Phys.~Lett. B {\bf 500}, 47 (2001).
\bibitem{Strauch03} S. Strauch \emph{ et al.}, \emph{ Phys.~Rev.~Lett.} {\bf 91}, 052301 (2003).
\bibitem{PR03104} Jefferson Lab Experiment E03-104, R.~Ent,
R.~Ransome, S.~Strauch, and P.~Ulmer, co-spokespersons.
\bibitem{Punjabi05} V.~Punjabi \emph{ et al.}, \emph{ Phys.~Rev.~C} {\bf 71}, 055202 (2005).
\bibitem{Ud98} J.M.~Udias \emph{ et al.},
        \emph{ Phys.~Rev.~Lett.} {\bf 83}, 5451 (1999);
        J.A.~Caballero, T.W. Donnelly, E. Moya de Guerra, and
        J.M. Udias, \emph{ Nucl.~Phys. }{\bf A632}, 323 (1998);
        J.M. Udias and J.R. Vignote, \emph{ Phys.~Rev.~C} {\bf 62}, 034302 (2000).
\bibitem{Forest83} T. de Forest, \emph{ Nucl.~Phys.~A} {\bf 392}, 232 (1983).
\bibitem{McNeil83} J.A.~McNeil, L.~Ray and S.J.~Wallace, \emph{ Phys.~Rev.~C}
{\bf 27}, 2123 (1983).
\bibitem{Horowitz} C.J. Horowitz, \emph{ Phys.~Rev.~C} {\bf 31} (1985) 1340; D.P. Murdock
and C.J. Horowitz, \emph{ Phys.~Rev.~C} {\bf 35} (1987) 1442.
\bibitem{meucci} A. Meucci, C. Giusti, and F.D. Pacati, \emph{ Phys.~Rev.~C} {\bf 66}, 034610 (2002).
\bibitem{Woo98} R.J. Woo \emph{ et al.}, \emph{ Phys.~Rev.~Lett.} {\bf 80}, 456 (1998).
\bibitem{Gao00} J. Gao \emph{ et al.}, \emph{ Phys.~Rev.~Lett.}. {\bf 84}, 3265 (2000).
\bibitem{Lava05} P. Lava \emph{ et al.}, \emph{ Phys.~Rev.~C} {\bf 71}, 014605 (2005).
\bibitem{Schiavilla05} R.~Schiavilla \emph{ et al.}, \emph{ Phys.~Rev.~Lett.} {\bf 94}, 072303 (2005).
\end{thebibliography}
\end{document}